\documentclass[12pt,epsf]{article}
\usepackage{graphicx}
\usepackage{epsfig}
\begin{document}

\def\a{\alpha}
\def\b{\beta}
\def\c{\varepsilon}
\def\d{\delta}
\def\e{\epsilon}
\def\f{\phi}
\def\g{\gamma}
\def\h{\theta}
\def\k{\kappa}
\def\l{\lambda}
\def\m{\mu}
\def\n{\nu}
\def\p{\psi}
\def\q{\partial}
\def\r{\rho}
\def\s{\sigma}
\def\t{\tau}
\def\u{\upsilon}
\def\v{\varphi}
\def\w{\omega}
\def\x{\xi}
\def\y{\eta}
\def\z{\zeta}
\def\D{\Delta}
\def\G{\Gamma}
\def\H{\Theta}
\def\L{\Lambda}
\def\F{\Phi}
\def\P{\Psi}
\def\S{\Sigma}

\def\o{\over}
\def\beq{\begin{eqnarray}}
\def\eeq{\end{eqnarray}}
\newcommand{\lsim}{\raisebox{0.6mm}{$\, <$} \hspace{-3.0mm}\raisebox{-1.5mm}{\em $\sim \,$}}
\newcommand{\gsim}{\raisebox{0.6mm}{$\, >$} \hspace{-3.0mm}\raisebox{-1.5mm}{\em $\sim \,$}}

\newcommand{\vev}[1]{ \left\langle {#1} \right\rangle }
\newcommand{\bra}[1]{ \langle {#1} | }
\newcommand{\ket}[1]{ | {#1} \rangle }
\newcommand{\EV}{ {\rm eV} }
\newcommand{\KEV}{ {\rm keV} }
\newcommand{\MEV}{ {\rm MeV} }
\newcommand{\GEV}{ {\rm GeV} }
\newcommand{\TEV}{ {\rm TeV} }
\def\diag{\mathop{\rm diag}\nolimits}
\def\Spin{\mathop{\rm Spin}}
\def\SO{\mathop{\rm SO}}
\def\O{\mathop{\rm O}}
\def\SU{\mathop{\rm SU}}
\def\U{\mathop{\rm U}}
\def\Sp{\mathop{\rm Sp}}
\def\SL{\mathop{\rm SL}}
\def\tr{\mathop{\rm tr}}

\def\IJMP{Int.~J.~Mod.~Phys. }
\def\MPL{Mod.~Phys.~Lett. }
\def\NP{Nucl.~Phys. }
\def\PL{Phys.~Lett. }
\def\PR{Phys.~Rev. }
\def\PRL{Phys.~Rev.~Lett. }
\def\PTP{Prog.~Theor.~Phys. }
\def\ZP{Z.~Phys. }


\baselineskip 0.7cm

\begin{titlepage}

\begin{flushright}
UT-08-09
\\
IPMU-08-0023
\end{flushright}

\vskip 1.35cm
\begin{center}
{\large \bf
   Strongly Interacting Gauge Mediation at the LHC
}
\vskip 1.2cm
Koichi Hamaguchi$^{1,2}$, Eita Nakamura$^1$, Satoshi Shirai$^1$ and T. T. Yanagida$^{1,2}$
\vskip 0.4cm

{\it $^1$  Department of Physics, University of Tokyo,\\
     Tokyo 113-0033, Japan\\
$^2$ Institute for the Physics and Mathematics of the Universe, 
University of Tokyo,\\ Chiba 277-8568, Japan}

\vskip 1.5cm

\abstract{
Strongly interacting gauge mediation (SIGM) of supersymmetry breaking is
very attractive, since it naturally predicts a light gravitino of mass $\lsim{\cal O}(10)$ eV, 
which causes no cosmological problem. 
We discuss various signatures of the SIGM in the early stage
(a low integrated luminosity period) of  the LHC experiments.
We show, in particular, a possible way 
to discriminate it from the conventional gauge mediation 
by counting the number of high $P_{\rm T}$ leptons.
}
\end{center}
\end{titlepage}

\setcounter{page}{2}

\section{Introduction}
\label{sec:1}
Strongly interacting gauge mediation (SIGM) of supersymmetry (SUSY) breaking~\cite{IY,INY}
is very attractive, since it naturally predicts
a light gravitino of mass $\lsim {\cal O}(10)$ eV and such a light gravitino
is free from all cosmological problems \cite{G1,G2,G3,G4}.
The SIGM predicts most likely a relatively light gluino compared with the conventional 
gauge mediation~\cite{Giudice:1998bp} and hence it may have more chance to be tested 
at the LHC.\footnote{\label{foot:1}
Note that the gluino pair production cross section is a steeply falling function of 
the gluino mass~\cite{Beenakker:1996ch}. 
In the conventional gauge mediation the gluino mass is comparable to the squark mass, while
the squark mass should be larger than approximately 1 TeV in order to satisfy the mass bound
on the lightest Higgs particle ($m_{h^0}>114.4$ GeV~\cite{LEPHiggsbounds}).
(The mass of SM-like Higgs receives a radiative correction from quark-squark loop diagrams.)
On the other hand, in the SIGM,
squarks are predicted much heavier than the gluino and hence the gluino mass can be 
in the range below 1 TeV, satisfying 
the Higgs mass bound.}
Therefore, an early SUSY discovery at the LHC with multiple jets and missing transverse momentum $P_{\rm T}$
may already indicate the SIGM. 

Furthermore, the next-to-lightest SUSY particle (NLSP) is the bino-like neutralino 
except for a very special case, which dominantly decays 
into the gravitino emitting a high energy photon. 
Hence, the SUSY events of multiple jets plus missing $P_{\rm T}$ are accompanied with 
two high $P_{\rm T}$ photons. 
Therefore, the early discovery of events with multiple jets $+$ missing $P_{\rm T}$ + two photons provides us 
with a crucial test of the SIGM.

In this letter, we discuss further tests of the SIGM at the LHC. For definiteness, we adopt an SIGM model
to calculate the spectrum of SUSY particles. We find that the gluino becomes lighter than
the wino in a large region of parameter space. 
We find that, as a consequence, the production of high $P_{\rm T}$ leptons 
is surprisingly suppressed at the LHC, contrary to the 
folklore of the generic SUSY phenomenology.

\section{A strongly interacting gauge mediation model}

The present model is an extension of the SUSY-breaking SU(5)$_{\rm hid}$ model
\cite{ADS} which has quark multiplets $Q$ transforming as
${\bf{5^*}} + {\bf 10}$ under the gauge group SU(5)$_{\rm hid}$.
We introduce two pairs of massive messengers $P_d +{\bar P_d}$ and $P_{\ell} + {\bar P_{\ell}}$ which
form ${\bf 5} + {\bf 5^*}$ of the standard-model (SM) SU(5)$_{\rm GUT}$.
A crucial assumption of the present SIGM model is that they also belong to ${\bf 5} + {\bf 5^*}$ of the
strongly interacting SU(5)$_{\rm hid}$ \cite{INY} and hence the messengers take part in the strong interactions
in the hidden sector. (See Table~\ref{table1}.)\footnote{We assume that the SM gauge symmetry is not broken by the strong dynamics.}

\begin{table}[t!]
\begin{center}
\begin{tabular}{c|c|c}
 & SU(5)$_{\rm GUT} \supset$ SM & SU(5)$_{\rm hid}$
 \\ \hline
 $Q$ & ${\bf 1}$ & ${\bf{5^*}} + {\bf 10}$
 \\
 $P_d + P_{\ell}$ & ${\bf 5}$ & ${\bf 5^*}$
 \\
 ${\bar P_d} + {\bar P_\ell}$ & ${\bf 5^*}$ & ${\bf 5}$
\\ \hline
 \end{tabular}
 \end{center}
\caption{The matter content of the SUSY breaking and messenger sectors.}
\label{table1}
\end{table}

We assume that the gauge coupling of the SU(5)$_{\rm hid}$ is strong at the messenger mass scale,
since otherwise we may have a split SUSY spectrum with very heavy scalars \cite{INY}. 
Unfortunately, we can not provide a
precise prediction of the masses of SUSY particles since the messenger particles are in the strong
interactions when they decouple and the integration of the messengers is uncalculable. 
Therefore, we adopt in this letter the naive dimensional
analysis (NDA) \cite{Luty} to estimate the masses of SUSY particles. 


To perform the NDA we assume, for simplicity,  that the SUSY is broken mostly by 
a composite state $\Phi_S$ which may be a bound state of the quarks $Q$ (${\bf 5^*} + {\bf 10}$).\footnote{The $\Phi_S$ may also contain covariant derivatives.}
The $\Phi_S$ is assumed to have both non-vanishing $A$ and $F$ terms to represent both  
the $R$ symmetry and SUSY breakings, which are known to occur by the strong dynamics \cite{ADS}.\footnote{The $R$ symmetry breaking results in an $R$-axion~\cite{Raxion}, but it is marginally consistent with the constraints (cf. Ref.~\cite{HallWatari}).}

The integration of the strongly interacting SU(5)$_{\rm hid}$ sector induces a 
low-energy effective K\"ahler potential. 
The relevant term for the scalar masses in the SUSY standard model (SSM) is given by
\beq K\simeq \frac{-1}{16\pi^2}g_{\rm SM}^4 \phi_{\rm SM}^\dagger \phi_{\rm SM}\frac{1}{16\pi^2}\frac{(g\Phi_S^\dagger)(g\Phi_S)}
{M_{d/\ell}^2}, \label{eq:s_kahler}
\eeq
where $\phi_{\rm SM}$ denote the SM superfields, $g_{\rm SM}$ are the SM gauge couplings, and 
$g$ is a constant of ${\cal O}(4\pi)$ representing the strong dynamics~\cite{Luty}.
$M_{d/\ell}$ are the effective masses of hadrons
$\Phi_{d/\ell}$
which consist of 
at least 
one messenger quark $P_{d/\ell}$ and SUSY-breaking quarks $Q$. 
Here, contributions of higher dimensional terms are ignored, for simplicity.
Eq.~(\ref{eq:s_kahler}) results in 
\beq m_\varphi^2 \simeq \Big(\frac{\alpha_{\rm SM}}{4\pi}\Big)^2\frac{|g\langle F_S\rangle|^2}{M_{d/\ell}^2}. \eeq
On the other hand, the K\"ahler potential relevant for the gaugino masses in the SSM is given by
\beq K\simeq g_{\rm SM}^2{\rm Tr}[W_{\rm SM}W_{\rm SM}]\frac{1}{16\pi^2}\frac{(g\Phi_S^\dagger)
(g\Phi_S)(g\Phi_S^\dagger)(gD^2\Phi_S)}{M_{d/\ell}^6}+{\rm h.c.}, \label{eq:g_kahler}
\eeq
which leads to
\beq m_\lambda\simeq \frac{\alpha_{\rm SM}}{4\pi}\frac{|g\langle F_S\rangle|^2g\langle \Phi_S^\dagger
\rangle g\langle F_S\rangle}{M_{d/\ell}^6}. \eeq
Notice that the gaugino masses arise not at ${\cal O}(F_S)$ but at ${\cal O}(F_S^3)$, since 
there is no direct coupling between the SUSY breaking fields and the messenger fields~\cite{INY,ArkaniHamed:1998kj}.

As for the masses of the messengers $P_d$ and $P_\ell$ we consider $m_d=m_\ell$ at the GUT scale.\footnote{$m_{d/\ell}$ are the masses of constituent messenger quarks $P_{d/\ell}$, while $M_{d/\ell}$ are the masses of composite hadrons $\Phi_{d/\ell}$.}
It is very important for our analysis that the SSM gauge interactions 
increase the value of $m_d$ more than $m_\ell$ at low energies. We find, by solving one-loop renormalization group (RG) equations for the messenger masses, 
$m_d\simeq 2.5\times m_\ell$ at the mass of the 
messenger $P_d$.\footnote{Here, we assume $m_d = {\cal O}(100)$~TeV to realize $m_{3/2}={\cal O}(1)$ eV 
and $m_{\rm gaugino}={\cal O}(100) $ GeV as we shall see later.}
 Below the threshold of the messenger $P_d$, only the messenger $P_\ell$
receives the mass renormalization from the strong SU(5)$_{\rm hid}$ interactions and 
hence the disparity in the messenger masses
becomes milder. Furthermore, the hadrons $\Phi_{d/\ell}$ contain the dynamical quarks $Q$ besides the messenger
quark $P_{d/\ell}$ and hence their mass ratio $M_d/M_\ell$ is smaller than the mass ratio of messenger quarks $m_d/m_\ell$.
Since the mass ratio is uncalculable due to the strong dynamics, we simply introduce one parameter $\kappa_1
=M_{d}/M_{\ell}$ and consider a region of $1\lsim \kappa_1 \lsim 2$.

Another important point is that the powers of $M_{d/\ell}$ are different between gauginos and scalars.
Assuming that $g\langle\Phi_S\rangle=\Lambda$ and $g\langle F_S\rangle=\Lambda^2$, we obtain
\beq m_\varphi^2\simeq \Big(\frac{\alpha_{\rm SM}}{4\pi}\Big)^2\frac{\Lambda^4}{M_{d/\ell}^2},
\qquad
m_\lambda\simeq \frac{\alpha_{\rm SM}}{4\pi}\frac{\Lambda^7}{M_{d/\ell}^6}. \label{eq:softmass} \eeq
Here, $\Lambda$ denotes the hadron mass scale. We introduce one more parameter $\kappa_2 = M_d/\Lambda$, which 
is of order unity and satisfies $\kappa_2 \gsim 1$. 
Note that $F_S=\Lambda^2/g
\simeq \Lambda^2/4\pi$ yields a relatively small gravitino mass compared with the 
case of $F_S\simeq\Lambda^2$ (i.e., $g\simeq 1$).

\section{Spectrum of the SUSY particles}

The low energy spectrum of the SSM particles can be obtained by using Eq.~(\ref{eq:softmass}) 
at the messenger mass scale 
and then RG-evolving their masses from the messenger mass scale down to the weak scale.
As discussed in the previous section, we introduce two parameters $\kappa_1=M_d/M_\ell$ and $\kappa_2=M_d/\Lambda$ 
to represent the uncertainties arising from the strong dynamics. We investigate the regions
\beq 1\le \kappa_1\le 2\qquad{\rm and}\qquad \kappa_2\ge 1. \eeq

The explicit mass formulae at the messenger mass scale are
\beq m_{\tilde{g}}=\frac{\alpha_3}{4\pi}\frac{\Lambda}{\kappa_2^6},\quad
m_{\widetilde{W}}=\frac{\alpha_2}{4\pi}\frac{\Lambda}{\kappa_2^6}\kappa_1^6,\quad
m_{\widetilde{B}}=\frac{\alpha_1}{4\pi}\frac{\Lambda}{\kappa_2^6}\bigg[\frac{1}{3}+\frac{1}{2}\kappa_1^6\bigg] \eeq
for the gauginos and
\beq m_\varphi^2=\frac{\Lambda^2}{\kappa_2^2}\bigg\{C_{{\rm SU(3)}_{C}}^\varphi\Big(\frac{\alpha_3}{4\pi}\Big)^2
+C_{{\rm SU(2)}_{L}}^\varphi\Big(\frac{\alpha_2}{4\pi}\Big)^2\kappa_1^2+C_{{\rm U(1)}_Y}^\varphi
\Big(\frac{\alpha_1}{4\pi}\Big)^2\bigg[\frac{1}{3}+\frac{1}{2}\kappa_1^2\bigg]\bigg\} \label{eq:scalarmass} \eeq
for the scalar particles, where $\alpha_i = g_i^2/4\pi$ are the SM gauge couplings 
and $C_{{\rm SU(3)}_{C}}^\varphi$, $C_{{\rm SU(2)}_{L}}^\varphi$ 
and $C_{{\rm U(1)}_Y}^\varphi$ are the quadratic Casimir invariants of
the corresponding gauge groups for the field $\varphi$.\footnote{
In our normalization of hypercharge, $\alpha_1=\frac{5}{3}\frac{g'^2}{4\pi}$ with $g'\cos\theta_W=e$
and 
$C_{{\rm U(1)}_Y}^\varphi=\frac{3}{5}\big(Q_{\rm em}^\varphi-T^{\varphi,3}_{\rm SU(2)}\big)^2$.}
We should mention as a reminder that there exist ${\cal O}(1)$ uncertainties in these formulae.

Before looking at the numerical results, we note some general features of the mass spectrum which can be
read from the above formulae. 
The parameter $\kappa_2$ dictates the hierarchy between the gaugino masses and the scalar masses.
As the value of $\kappa_2$ increases, the scalars become heavier than the gauginos. 
The parameter $\kappa_1$ determines the relative 
mass relations between SU(3)$_{C}$ charged particles ($\tilde{g}$ and $\tilde{q}$)
 and the SU(3)$_{C}$ neutral particles ($\tilde{\ell}$, $\tilde{\chi}$ and $H_{u,d}$). 
As the value of $\kappa_1$ increases, SU(3)$_{C}$ neutral particles become heavier. 
In the limit $\kappa_1,\kappa_2\to1$, we recover the GUT relation for the gaugino masses and $m_{\rm scalar}\sim m_{\rm gaugino}$
as in the conventional gauge mediation model. 
$\Lambda$ provides the overall scale for the soft SUSY breaking masses.

As remarked earlier, the gravitino is very light in our model. Its mass is given by
\beq
 m_{3/2}=\frac{\langle F_S\rangle}{\sqrt{3}M_P}\simeq\frac{\Lambda^2}
{4\pi\sqrt{3}M_P}, 
\eeq
where $M_P=2.44\times10^{18}$ GeV is the reduced Planck mass.
We see that $m_{3/2}<10$ eV corresponds to $\Lambda\lsim 730$ TeV.

An important issue is the occurrence of the electroweak (EW) symmetry breaking.\footnote{We do not discuss the origin of the $\mu$- and $B$-terms (the $\mu$-problem) in the present model and take the $\tan\beta$ as a free parameter.}
In our model,
the Higgs scalars can be relatively heavy and accordingly, we must check
whether the squared Higgs masses become so small at the weak scale that
the EW symmetry breaking occurs. In fact, as the value of $\kappa_1$ increases, the soft SUSY breaking 
Higgs masses squared 
$m_{H_{u,d}}^2$ become larger and there is an upper bound on $\kappa_1$ (for each value of $\kappa_2$) above which 
no EW symmetry breaking occurs.

\begin{figure}[htbp]
\begin{tabular}{c}
\begin{minipage}{\hsize}
\begin{center}
\epsfig{file=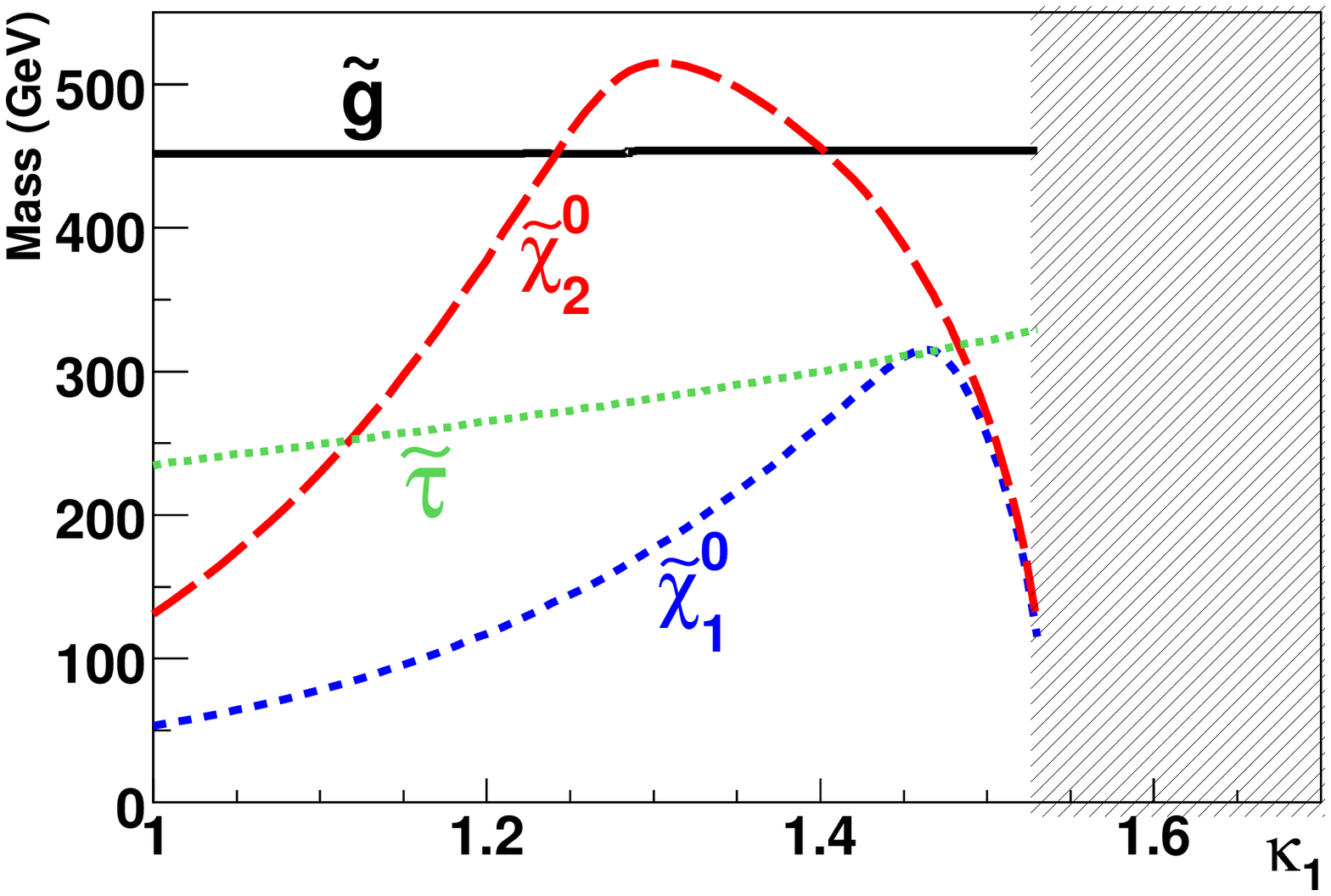 ,scale=.5,clip}
\caption{Mass spectrum for  $\Lambda=280$ TeV and $\kappa_2=1.35$. 
Here, scalar quark masses are approximately 1.5 TeV. 
In the shaded region, EW breaking dose not occur.
}
\label{fig:mass280}
\end{center}
\end{minipage}\\
\begin{minipage}{\hsize}
\begin{center}
\epsfig{file=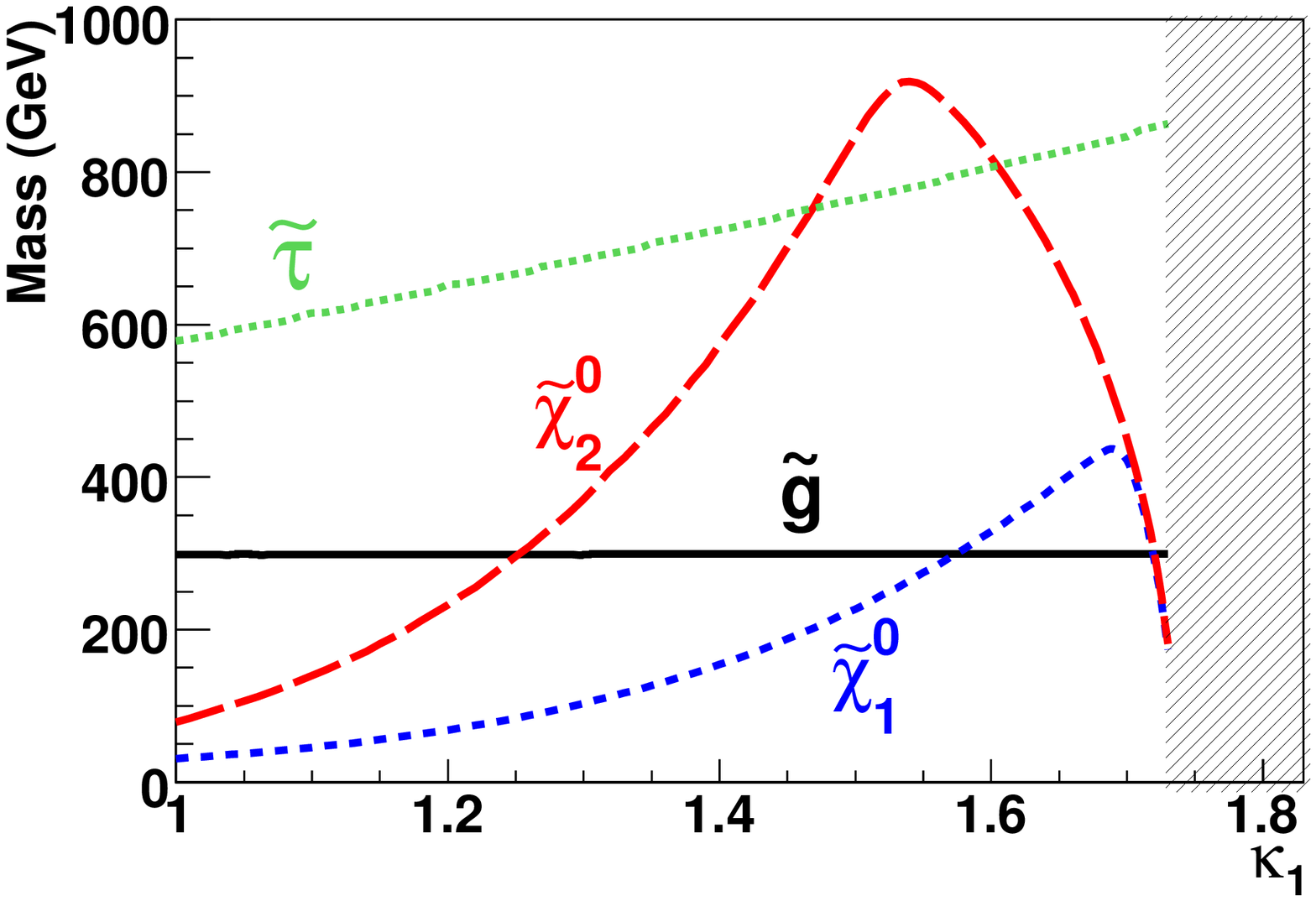 ,scale=.5,clip}
\caption{Mass spectrum for  $\Lambda=900$ TeV and $\kappa_2=1.8$. 
Here, scalar quark masses are approximately 3.5 TeV. 
In the shaded region, EW breaking dose not occur.
}
\label{fig:mass900}
\end{center}
\end{minipage}
\end{tabular}
\end{figure}

Now we present the numerical results. We calculate the SUSY-particle masses by using 
the one-loop RG equations for the SSM parameters.
In Figs.~\ref{fig:mass280} and \ref{fig:mass900}, we show the gluino mass ($m_{\tilde{g}}$), 
the first two lightest neutralino masses ($m_{\tilde{\chi}_1^0},m_{\tilde{\chi}_2^0}$) 
and the lighter stau mass $(m_{\tilde{\tau}_1})$ as functions of $\kappa_1$. 
Two examples for $(\kappa_2, \Lambda) = (1.35, 280~{\rm TeV})$ and
$(1.8, 900~{\rm TeV})$  are shown in Figs.~\ref{fig:mass280} and \ref{fig:mass900}, respectively.
Here, $\tan\beta=10$ in both cases. 
The upper bound of $\kappa_1$ in each figure corresponds to the point where $\mu=0$ above which  EW symmetry breaking 
does not occur. 

The general features of the mass spectrum discussed above can be seen in those examples.
The NLSP is the lightest neutralino in most of the parameter region.\footnote{
The lightest neutralino is bino-like 
and the second lightest neutralino is wino-like in a large region of the parameter space where $\kappa_1$ 
is not near the upper bound value.
As $\kappa_1$ approaches the upper bound, $\mu$ becomes much smaller than $m_{\widetilde{B}}$ and 
$m_{\widetilde{W}}$, and the higgsino components dominate in the first two lightest neutralino,
whose masses approach zero when $\kappa_1$ is near the upper bound.}
The GUT relation among the gaugino masses, which holds in the conventional gauge mediation models, is violated for
$\kappa_1>1$. In particular, for $\kappa_1\gsim 1.2$, $m_{\widetilde{W}}$ becomes larger than $m_{\tilde{g}}$.
Note also that the gluino becomes the NLSP for $\kappa_1\gsim 1.6$ in Fig.~\ref{fig:mass900}. 
The scalars are heavier compared with in the conventional gauge mediation for $\kappa_2>1$. The stau is the lightest among the scalar particles, whose mass is also shown in Figs.~\ref{fig:mass280} and \ref{fig:mass900}.

\section{Test for $m_{\widetilde{W}} > m_{\widetilde{g}}$ at the LHC}
If the SIGM is the case and the neutralino is the NLSP, its experimental signatures are high $P_{\mathrm T}$ photons,
high $P_{\mathrm T}$ multiple jets and large
 missing $P_{\mathrm T}$.
However, the conventional minimal gauge mediated SUSY breaking (mGMSB) models with ultralight gravitino LSP and neutralino NLSP will also have similar signatures.
Here, we discuss possible ways to discriminate between those two models.

We have seen that the SIGM predicts heavier scalar particles 
and violation of GUT relation for the gaugino masses.
Therefore, we can straightforwardly discriminate the SIGM models 
from the mGMSB ones by measuring the SUSY particles' masses. 
However, it may require a large integrated luminosity.
In the following, we propose another way to discriminate the SIGM at an earlier stage of the LHC experiments.

At the LHC, if the SIGM is the case, SUSY particles are mainly produced through: 
\begin{equation}
pp\rightarrow \tilde{g}+\tilde{g}+X.
\end{equation}
In the case that $m_{\widetilde{W}} > m_{\widetilde{g}}$,
the gluinos dominantly decay into $\tilde{B}+q+\bar{q}$, $\tilde{B} + g$ and $\tilde{B}+t+\bar{t}$ if not kinematically forbidden.
Then, the produced binos $\tilde{B}$ dominantly decay into $\gamma+ \tilde{G}_{3/2}$, and $Z^0+ \tilde{G}_{3/2}$ if not kinematically forbidden ($\tilde{G}_{3/2}$ denotes the gravitino). 
Thus if $m_{\widetilde{W}} > m_{\widetilde{g}}$ is the case, there are no high $P_{\mathrm T}$ leptons except for ones which come from $Z^0$'s and $t$'s decays. 

On the other hand, in the case of mGMSB, many lepton production channels exist
since the wino (and sleptons) are lighter than the gluino.
Therefore, we can distinguish between mGMSB and SIGM by counting the number of  high $P_{\rm T}$ leptons.
We consider two examples, by taking an SIGM ($m_{3/2}=10$ eV, $\kappa_1 = 1.35$, $\kappa_2 =1.5$, $\tan\beta=10$)
and an mGMSB ($F/M_{\rm mess} = 80$ TeV, $M_{\rm mess} = 160$ TeV, $N_5 =1$, $\tan\beta=10$).
The mass spectrums are shown in Figs.~\ref{fig:sigm} and \ref{fig:mgmsb}.\footnote{
This mGMSB example has a light gluino, but it does not satisfy the Higgs mass bound. (See the discussion in Sec.~\ref{sec:1}.)
We take this model point just as a demonstration, for a comparison to the SIGM.
}
These spectrums are calculated by ISAJET7.72~ \cite{ISAJET}.
To simulate LHC signatures for these models, we use programs Herwig 6.5~\cite{HERWIG6510} and AcerDET-1.0~\cite{RichterWas:2002ch}.

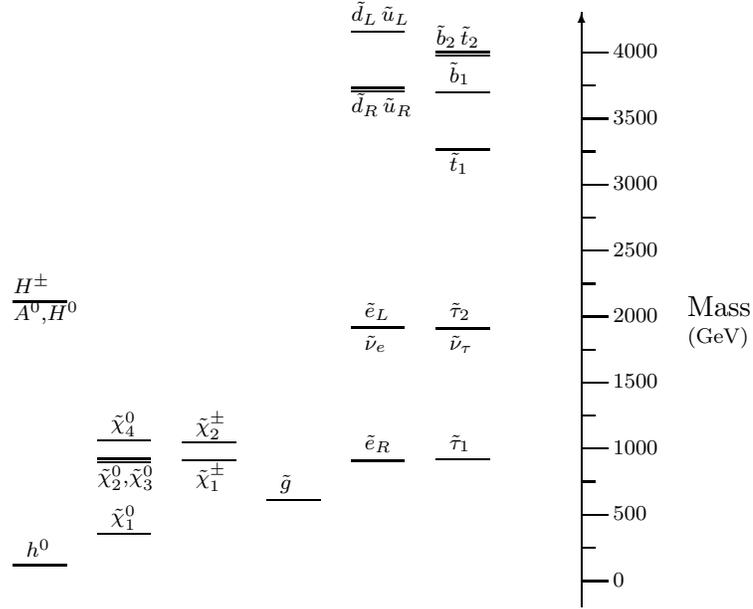
\begin{figure}
\begin{center}

\scalebox{1}{
\begin{picture}(220,200.12)(0,0)
\put(5,5.9945){\line(1,0){20}}
\put(10,8.9945){$\scriptstyle h^0$}
\put(5,105.87){\line(1,0){20}}
\put(5,105.465){\line(1,0){20}}
\put(5,105.941){\line(1,0){20}}
 \put(5,108.941){$\scriptstyle H^{\pm}$}
 \put(5,98.8695){$\scriptstyle A^0,H^0$}
\put(37,17.793){\line(1,0){20}}
\put(42.,21.793){$\scriptstyle {\tilde{\chi}^0_1}$}
\put(37,44.9685){\line(1,0){20}}
\put(37.,36.9685){$\scriptstyle {\tilde{\chi}^0_2,\tilde{\chi}^0_3}$}
\put(37,46.2375){\line(1,0){20}}
\put(37,53.1705){\line(1,0){20}}
\put(42.,57.1705){$\scriptstyle {\tilde{\chi}^0_4}$}
\put(69,45.7355){\line(1,0){20}}
\put(74.,36.7355){$\scriptstyle {\tilde{\chi}^{\pm}}_1$}
\put(69,52.313){\line(1,0){20}}
\put(74.,56.313){$\scriptstyle {\tilde{\chi}^{\pm}_2}$}
\put(101,30.63){\line(1,0){20}}
\put(106,34.63){$\scriptstyle {\tilde g}$}
\put(133,207.863){\line(1,0){20}}
\put(133,207.824){\line(1,0){20}}
\put(133,211.863){$\scriptstyle \tilde d_L\,\tilde u_L$}
\put(133,185.325){\line(1,0){20}}
\put(133,186.641){\line(1,0){20}}
\put(133,177.325){$\scriptstyle \tilde d_R\,\tilde u_R$}
\put(133,95.98){\line(1,0){20}}
\put(133,95.584){\line(1,0){20}}
\put(138,87.584){$\scriptstyle \tilde \nu_e$}
\put(138,99.98){$\scriptstyle \tilde e_L$}
\put(133,45.487){\line(1,0){20}}
\put(138,50.0){$\scriptstyle \tilde e_R $}
\put(165,163.232){\line(1,0){20}}
\put(170,155.232){$\scriptstyle \tilde t_1$}
\put(165,200.12){\line(1,0){20}}
\put(165,184.816){\line(1,0){20}}
\put(170,188.816){$\scriptstyle \tilde b_1$}
\put(165,198.896){\line(1,0){20}}
\put(165,202.896){$\scriptstyle \tilde b_2\,\tilde t_2$}
\put(165,45.9365){\line(1,0){20}}
\put(170,49.9365){$\scriptstyle \tilde \tau_1$}
\put(165,95.7455){\line(1,0){20}}
\put(170,99.7455){$\scriptstyle \tilde \tau_2$}
\put(165,95.4025){\line(1,0){20}}
\put(170,87.4025){$\scriptstyle \tilde \nu_{\tau}$}
\put(260,100.06){\small Mass }
\put(260,90.06){$\scriptstyle {\rm (GeV)}$ }
\put(220,-10){\vector(0,1){225.12}}
\put(220,0){\line(1,0){10}}
\put(232,-2){$\scriptstyle 0$}
\put(220,25){\line(1,0){10}}
\put(232,23){$\scriptstyle 500$}
\put(220,50){\line(1,0){10}}
\put(232,48){$\scriptstyle 1000$}
\put(220,75){\line(1,0){10}}
\put(232,73){$\scriptstyle 1500$}
\put(220,100){\line(1,0){10}}
\put(232,98){$\scriptstyle 2000$}
\put(220,125){\line(1,0){10}}
\put(232,123){$\scriptstyle 2500$}
\put(220,150){\line(1,0){10}}
\put(232,148){$\scriptstyle 3000$}
\put(220,175){\line(1,0){10}}
\put(232,173){$\scriptstyle 3500$}
\put(220,200){\line(1,0){10}}
\put(232,198){$\scriptstyle 4000$}
\put(220,0){\line(1,0){5}}
\put(220,12.5){\line(1,0){5}}
\put(220,25){\line(1,0){5}}
\put(220,37.5){\line(1,0){5}}
\put(220,50){\line(1,0){5}}
\put(220,62.5){\line(1,0){5}}
\put(220,75){\line(1,0){5}}
\put(220,87.5){\line(1,0){5}}
\put(220,100){\line(1,0){5}}
\put(220,112.5){\line(1,0){5}}
\put(220,125){\line(1,0){5}}
\put(220,137.5){\line(1,0){5}}
\put(220,150){\line(1,0){5}}
\put(220,162.5){\line(1,0){5}}
\put(220,175){\line(1,0){5}}
\put(220,187.5){\line(1,0){5}}
\end{picture} }
\caption[]{SIGM mass spectrum for $m_{3/2}=10$ eV, $\kappa_1 = 1.35$, $\kappa_2 =1.5$ and $\tan\beta=10$.}
\label{fig:sigm}
\end{center}
\end{figure}
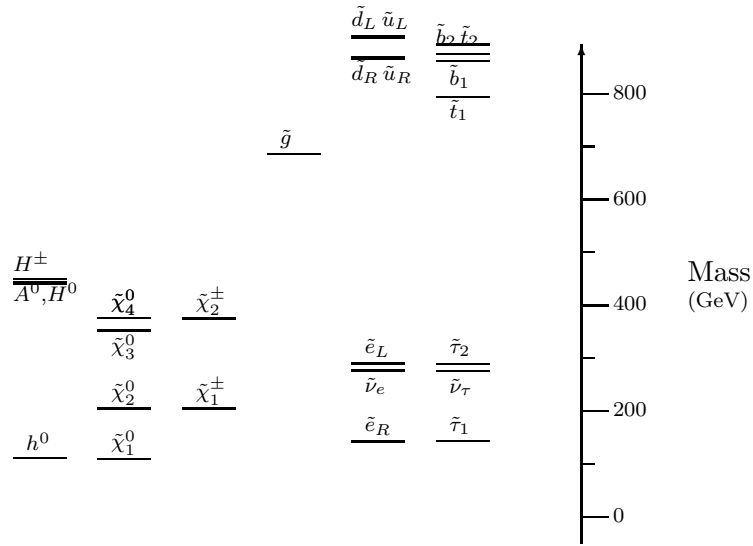
\begin{figure}
 \begin{center}
\scalebox{1}{
\begin{picture}(220,178.694)(0,0)
\put(5,22.086){\line(1,0){20}}
\put(10,25.086){$\scriptstyle h^0$}
\put(5,88.724){\line(1,0){20}}
\put(5,88.3){\line(1,0){20}}
\put(5,90.056){\line(1,0){20}}
 \put(5,93.056){$\scriptstyle H^{\pm}$}
 \put(5,81.724){$\scriptstyle A^0,H^0$}
\put(37,21.946){\line(1,0){20}}
\put(42.,25.946){$\scriptstyle {\tilde{\chi}^0_1}$}
\put(37,40.916){\line(1,0){20}}
\put(42.,44.916){$\scriptstyle {\tilde{\chi}^0_2}$}
\put(37,70.4){\line(1,0){20}}
\put(42.,62.4){$\scriptstyle {\tilde{\chi}^0_3}$}
\put(37,75.184){\line(1,0){20}}
\put(42.,79.184){$\scriptstyle {\tilde{\chi}^0_4}$}
\put(37,75.184){\line(1,0){20}}
\put(42.,79.184){$\scriptstyle {\tilde{\chi}^0_4}$}
\put(69,40.932){\line(1,0){20}}
\put(74.,44.932){$\scriptstyle {\tilde{\chi}^{\pm}}_1$}
\put(69,75.054){\line(1,0){20}}
\put(74.,79.054){$\scriptstyle {\tilde{\chi}^{\pm}_2}$}
\put(101,137.132){\line(1,0){20}}
\put(106,141.132){$\scriptstyle {\tilde g}$}
\put(133,181.904){\line(1,0){20}}
\put(133,181.2){\line(1,0){20}}
\put(133,185.904){$\scriptstyle \tilde d_L\,\tilde u_L$}
\put(133,173.42){\line(1,0){20}}
\put(133,173.926){\line(1,0){20}}
\put(133,165.42){$\scriptstyle \tilde d_R\,\tilde u_R$}
\put(133,57.892){\line(1,0){20}}
\put(133,55.25){\line(1,0){20}}
\put(138,47.25){$\scriptstyle \tilde \nu_e$}
\put(138,61.892){$\scriptstyle \tilde e_L$}
\put(133,28.346){\line(1,0){20}}
\put(138,32.346){$\scriptstyle \tilde e_R $}
\put(165,158.98){\line(1,0){20}}
\put(170,150.98){$\scriptstyle \tilde t_1$}
\put(165,178.694){\line(1,0){20}}
\put(165,172.506){\line(1,0){20}}
\put(170,163.506){$\scriptstyle \tilde b_1$}
\put(165,175.098){\line(1,0){20}}
\put(165,179.098){$\scriptstyle \tilde b_2\,\tilde t_2$}
\put(165,28.526){\line(1,0){20}}
\put(170,32.526){$\scriptstyle \tilde \tau_1$}
\put(165,57.766){\line(1,0){20}}
\put(170,61.766){$\scriptstyle \tilde \tau_2$}
\put(165,54.968){\line(1,0){20}}
\put(170,46.968){$\scriptstyle \tilde \nu_{\tau}$}
\put(260,89.347){\small Mass }
\put(260,79.347){$\scriptstyle {\rm (GeV)}$ }
\put(220,-10){\vector(0,1){188.694}}
\put(220,0){\line(1,0){10}}
\put(232,-2){$\scriptstyle 0$}
\put(220,40){\line(1,0){10}}
\put(232,38){$\scriptstyle 200$}
\put(220,80){\line(1,0){10}}
\put(232,78){$\scriptstyle 400$}
\put(220,120){\line(1,0){10}}
\put(232,118){$\scriptstyle 600$}
\put(220,160){\line(1,0){10}}
\put(232,158){$\scriptstyle 800$}
\put(220,0){\line(1,0){5}}
\put(220,20){\line(1,0){5}}
\put(220,40){\line(1,0){5}}
\put(220,60){\line(1,0){5}}
\put(220,80){\line(1,0){5}}
\put(220,100){\line(1,0){5}}
\put(220,120){\line(1,0){5}}
\put(220,140){\line(1,0){5}}
\end{picture} }

\caption[]{mGMSB mass spectrum $F/M_{\rm mess} = 80$ TeV, $M_{\rm mess} = 160$ TeV, $N_5 =1$ and $\tan\beta=10$.}
\label{fig:mgmsb}
\end{center}
\end{figure}

We take the events cuts as follows:
\begin{itemize}
\item $\ge 4$ jets with $P_{\rm T}>50$ GeV and $P_{\rm T,1,2}>100$ GeV.
\item $\ge 2$ photons with $P_{\rm T}>10$ GeV and $P_{\rm T,1}>20$ GeV.
\item $M_{\rm eff}>500$ GeV, where
\begin{equation}
M_{\rm eff} = \sum_{\rm jets}^{4} P_{{\rm T}j} + P_{\rm T, miss}.
\end{equation} 
\item $P_{\rm T, miss}>0.2 M_{\rm eff}$.
\end{itemize}
Under these cuts, we see that the standard model backgrounds are almost negligible.

\begin{figure}[t]
\begin{center}
\epsfig{file=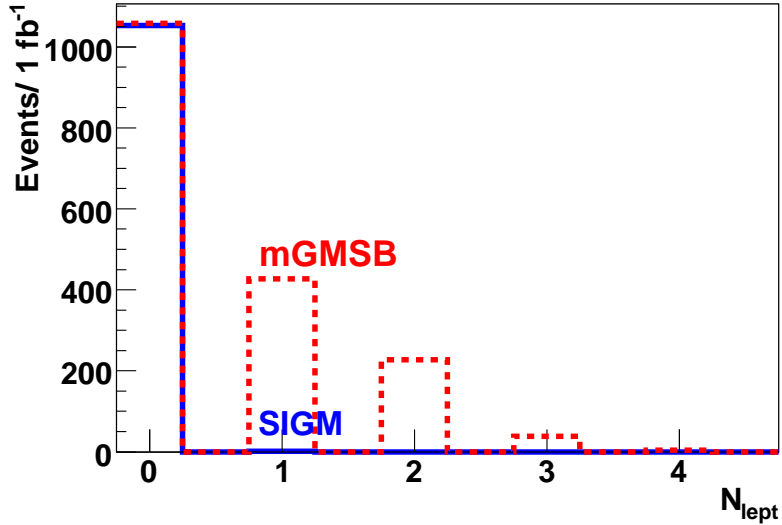 ,scale=.55,clip}
\caption{The lepton number distributions for mGMSB and SIGM.}
\label{fig:lepton}
\end{center}
\end{figure}
The number distributions for high $P_{\rm T}$ leptons (with $P_{\rm T}>20$ GeV) are shown in Fig.~\ref{fig:lepton}.
In Fig.~\ref{fig:lepton}, we can see there are very few leptons produced at the LHC in the SIGM. 
The photon cuts reduce the lepton production from $\tilde{B}$'s decay,
 and the decay $\tilde{g}\rightarrow \tilde{B}+t+\bar{t}$ is kinematically forbidden in the SIGM example.

\begin{figure}[htbp]
\begin{tabular}{c}
\begin{minipage}{\hsize}
\begin{center}
\epsfig{file=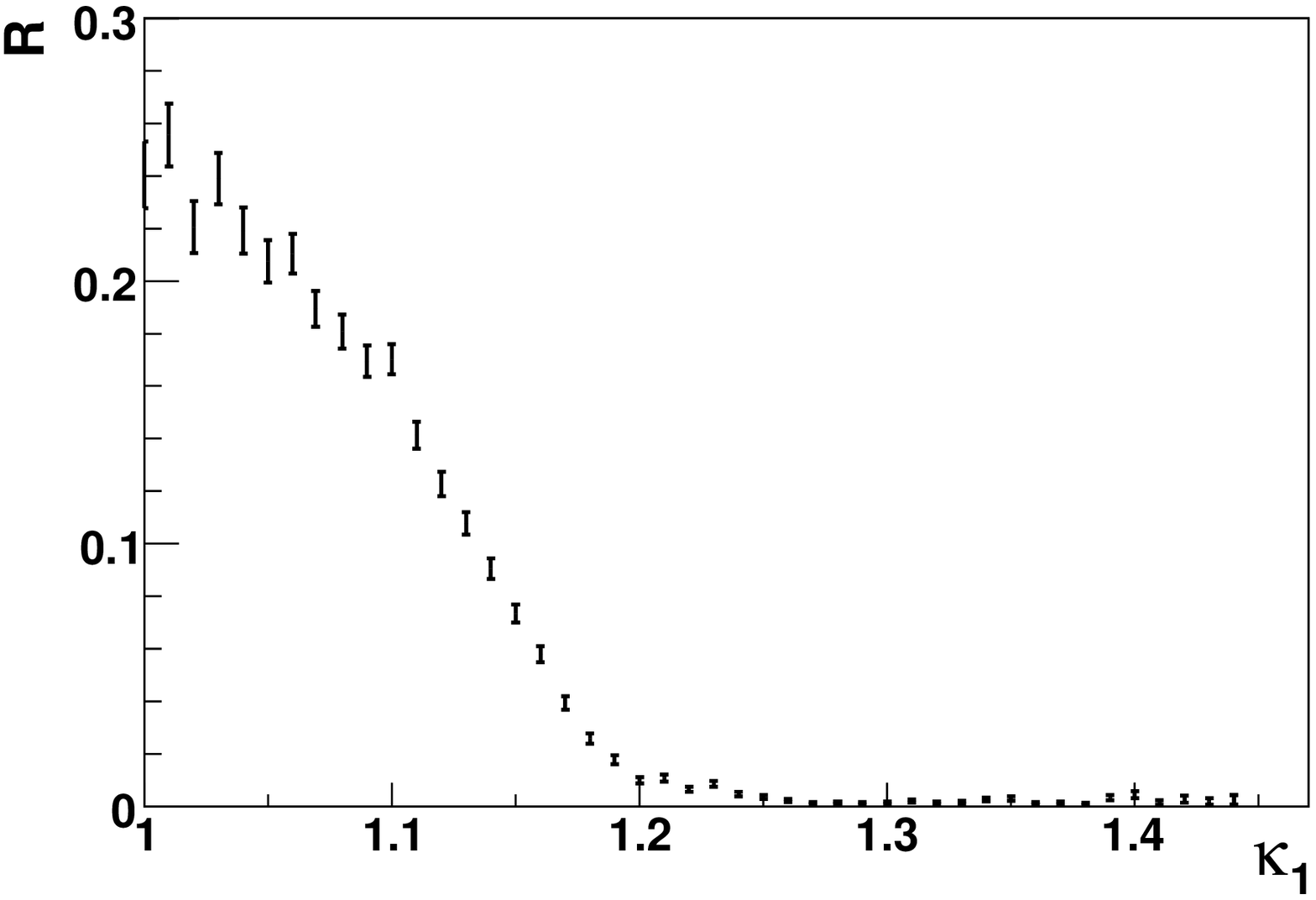 ,scale=.6,clip}
\caption{The relation of $\kappa_1$ to $R$.
SIGM parameters are the same as in Fig.~\ref{fig:sigm} except for $\kappa_1$.
Error bars represent only the statistical errors.
}
\label{fig:ratio}
\end{center}
\end{minipage}\\
\begin{minipage}{\hsize}
\begin{center}
\epsfig{file=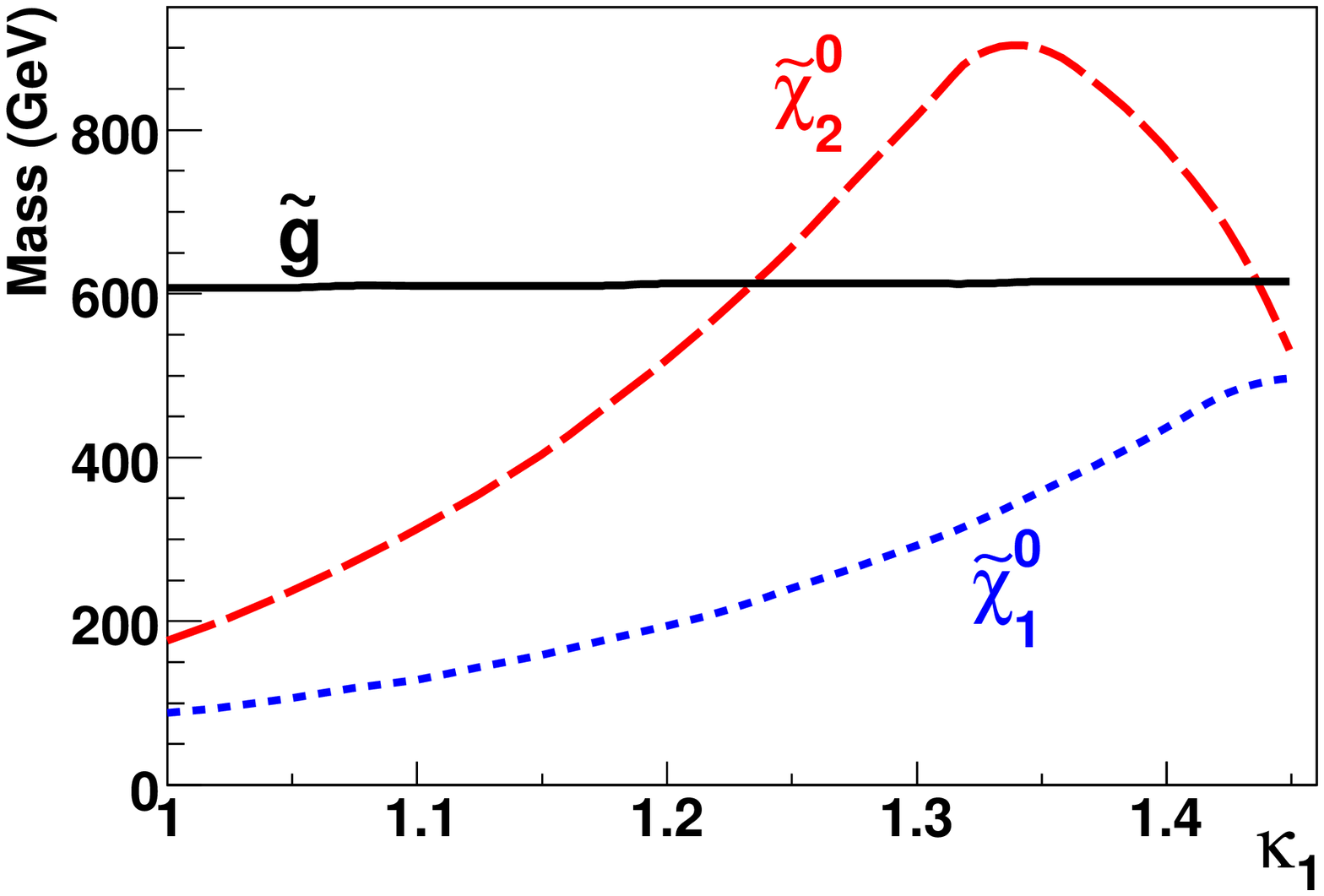 ,scale=.6,clip}
\caption{The relation of $\kappa_1$ to $m_{\tilde{g}}$ and $m_{\tilde{\chi}^0_{1,2}}$.}
\label{fig:massisajet}
\end{center}
\end{minipage}
\end{tabular}
\end{figure}

In general, larger $\kappa_1$ results in the smaller production 
rate for high $P_{\rm T}$ leptons.
To see the relation of $\kappa_1$ to the lepton production rate, 
we define $R$ as
\beq
R \equiv \frac{ {\rm \#~ of~ events ~after~ the 
~ cuts ~with~ at~ least~ one~ lepton}~ (P_{\rm T}>20 ~{\rm GeV})}
{\rm \# ~of~ all~ events~ after ~the~ cuts }.
\eeq
For the conventional mGMSB with the parameters given above, we obtain $R=0.40$.
In Figs.~\ref{fig:ratio} and \ref{fig:massisajet} we show the 
$\kappa_1$ dependence of $R$ and ($m_{\widetilde{g}}$, $m_{\tilde{\chi}^0_1}$, $m_{\tilde{\chi}^0_2}$) 
in the SIGM ($m_{3/2}=10$ eV, $\kappa_2 =1.5$ and $\tan\beta=10$), respectively.
To compute $R$, 20000 SUSY events are generated for each value of $\kappa_1$.
Fig.~\ref{fig:ratio} clearly shows that there are less lepton production for 
larger $\kappa_1$.
It is also confirmed that in the case $m_{\tilde{W}}\gsim m_{\tilde{g}}$ almost no leptons are produced.
Thus, we may distinguish between mGMSB and SIGM by counting the number of high $P_{\rm T}$ leptons.

\section{Discussion}
The strongly interacting gauge mediation (SIGM) predicts gaugino masses without the GUT relation, in particular
a relatively light gluino compared with a conventional gauge mediation. 
We have calculated the SUSY mass spectrum by taking some explicit examples, and shown that 
the SIGM with such a mass spectrum can be discriminated simply by counting the number of high $P_{\rm T}$ leptons.

If many events with two photons, multiple jets and missing energy are discovered at the LHC, it naturally points to 
gauge mediation models with a neutralino NLSP and the gravitino LSP. 
If the number of such events is large, which suggests the light gluino or squark, 
it may already indicate an unconventional gauge mediation mass spectrum even at an early stage of 
the LHC experiments.\footnote{Note that the gluino and squarks are heavy in the conventional GMSB models. See footnote~\ref{foot:1}.}
We can then test a peculiar mass hierarchy among gauginos by simply counting the number of high $P_{\rm T}$ leptons, as investigated in this letter.

We have also found that the gluino can even be the NLSP for some parameter region. In this case the main SUSY event signal will be two jets + missing energy. 
In such a case, it will be challenging to identify the LSP (gravitino) and the NLSP (gluino).

\section*{Acknowledgements}

This work was supported by World Premier International Center Initiative
(WPI Program), MEXT, Japan.
The work of TTY is supported in part by the Grant-in-Aid for Science
Research, Japan Society for the Promotion of Science, Japan (No.~1940270).
The work by KH is supported by JSPS (18840012).
The work of SS is supported in part by JSPS
Research Fellowships for Young Scientists.

\end{document}